\documentstyle[aps,epsf,12pt]{revtex}
\tightenlines
\begin{document}

\title{Robust Nuclear Observables and Constraints on Random Interactions}
 \author{Dimitri Kusnezov$^a$, N.V. Zamfir$^{b,c,d}$ and R.F. Casten$^b$}
\address{$^a$Center for Theoretical Physics, Sloane Physics Lab, Yale
  University, New Haven, CT\ 06520-8120\\
  $^b$Wright Nuclear Structure Laboratory, Yale University, New 
Haven, CT 06520-8120\\
$^c$  Clark University, Worcester, Massachusetts 01610\\
$^d$ National Institute for Physics and Nuclear Engineering,
 Bucharest-Magurele, Romania} \date{\today }
\maketitle

\begin{abstract}
 The predictions of the IBM two-body random ensemble are
compared to empirical results on
nuclei from Z=8 to 100.  Heretofore unrecognized but robust
empirical trends are identified and related both to the
distribution of valence nucleon numbers and to the need for and
applicability of specific, non-random interactions.  Applications to
expected trends in exotic nuclei are discussed.
\end{abstract}


Recently there has been renewed interest in the spectroscopic
properties of Hamiltonians with random
interactions~\cite{1,2,3,4,5,6,7}. Scrutiny has focused on nuclear
systems but applications exist in other many body quantal systems
such as quantum dots or metallic clusters.
  In the nuclear case, analyses have been carried out
in the frameworks of the Shell Model~\cite{1,2,3,5} and the
Interacting Boson Model (IBM)~\cite{4,5,6}.

These studies are extremely important because they explore the
origins of fundamental features of nuclear structure. The
upshot of these studies has been surprising recognition that some
of the most hallowed aspects of structure, such as $0^+$ ground
states for even-even nuclei, and even vibrational spectra, arise
with randomly distributed interactions.

It is the purpose of this Letter to confront calculations with
random and non-random interactions with robust experimental
results to study which pervasive features of nuclei arise from the
basis space and the generic type of interaction (e.g., 2-body) and
which depend on the details of the interaction. We will use the
insights developed in this analysis to better understand
structural evolution and its relation to shell structure, and to
project the behavior of exotic nuclei.

First, it is useful to succinctly summarize existing studies. In
Shell model \cite{1,2,3} and IBM \cite{4,5,6}
approaches for many-body systems, the majority
(typically $\sim$70\%) of calculated ground states have
$J^{\pi}=0^+$ even though $0^+$ states comprise a much smaller
fraction of the basis space than other angular momenta. While it
might be suspected that such a "pairing" property could emerge as
a consequence of time reversal invariance, it has been
shown~\cite{2} that this is not the case. Comparison of
ground state wave functions in nuclei differing by two
nucleons shows \cite{3} further evidence for a pairing relationship,
reflecting the generalized seniority
scheme~\cite{8}.  The analysis also extends to excited states.  In
the Shell Model, modest probabilities were found \cite{1} for
vibration-like spectra, defined by energy level ratios
$R_{4/2}\equiv{E(4^+_1)/E(2^+_1)}$ near 2.0. However it was not
possible to produce significant numbers of spectra with rotational
character, $R_{4/2}\sim{10/3}$.

In the IBM analyses~\cite{4,5,6} a similar preponderance of $0^+$
ground states was found.  Here, however, in contrast to the Shell
Model, the interactions in an s-d boson space lead rapidly, with
increasing boson number $N_B$, to large probabilities for both
vibrational [$R_{4/2}\sim{2.0}$] and rotational
[$R_{4/2}\sim{3.33}$] spectra. Indeed, the IBM treatments produced
no concentrations of $R_{4/2}$ values near $\it{any\ other}$
values, as noted in Refs. \cite{4,5,6}. These results are
reminiscent of empirical 
correlations~\cite{9,10} of $4^+_1$ and $2^+_1$ energies which
give evidence for anharmonic vibrator and rotor behavior.

These studies~\cite{1,2,3,4,5,6} are both surprising and
important, and potentially impact our understanding of the
origins of some of the most characteristic features of nuclear
spectra. Such features seem to 
emerge primarily from the nature of the basis space and the rank 
(2-body) of the interaction.

Despite all this recent work, there has been no explicit
comparison with experiment.  It is therefore one of the purposes
of the present Letter to do so, with data on level energies
spanning the entire
nuclear chart\cite{10a}. This is the first time that detailed data on the
low-lying nuclear spectra have been compared with
predictions obtained from random ensembles. In the process,
some previously unrecognized features of the data themselves will
be discerned, which are not in fact produced with random
interactions, giving us the kind of information needed to define
favored interactions.

To proceed, we use two IBM Hamiltonian ensembles, the full, most general,
Hamiltonian (denoted $H_{IBM}$) with two 1-body and seven 2-body
interactions, and the more ``focused" Hamiltonian
$H_{\epsilon\kappa}=\epsilon n_{d}-{\kappa}Q\cdot{Q}$ of the extended
consistent Q formalism (ECQF)~\cite{11,12}, where
$Q=s^\dagger\tilde d +d^\dagger s +\chi [d^\dagger\tilde
d]^{(2)}$. 
We use either uniform,
random $\chi\ \in\ [0,-\sqrt7/2]$ or fixed $\chi$, as specified in
each set of calculations below. The ensembles are defined by
choosing random Gaussian interaction strengths in the
Hamiltonians $H_{IBM}$ and $H_{\epsilon\kappa}$ as in
ref.\cite{4}. The full Hamiltonian gives results
essentially identical to those in Ref. [4], with peaks at
$R_{4/2}\sim{2.0}$ and 3.33, and growing probability for the
latter as $N_B$ increases. The focused ECQF Hamiltonian (with
random $\chi$ in the Q operator) gives similar but ``cleaner"
$R_{4/2}$ distributions.  Both results are shown in the inset to
Fig. 1. It is interesting to examine the correlations of
$E(4^+_1)$ vs. $E(2^+_1)$. Empirically, this
correlation has been shown~\cite{9,10} to be extremely robust,
exhibiting a bi-modal character with slopes of 2.0 and 3.33 in
collective nuclei.
Comparisons of these data with calculations using the full
H$_{IBM}$ for boson number N$_B$=16 are shown in Fig. 1.
The overall energy scale is fixed to the data by choosing the
width of the Gaussian distributions. Similar results are obtained
for other $N_B$.  

While the overall agreement is excellent, the
trends of Fig. 1 only show part of the story.  They do not easily
reveal the $\it{density}$ distribution along the trajectory, that
is, the probability distribution of nuclei as a $\it{function}$ of
$R_{4/2}$. And it is from this perspective that specific empirical
features appear that are not described by calculations unless the range
of interactions is constrained. To see this, consider Fig. 2(left)
which shows empirical $R_{4/2}$ values for the entire nuclear
chart from Z=8-100. In light nuclei, there is little evidence for
collective $R_{4/2}$ values. In the lowest medium mass region,
$28<Z<50$, however, a peak at $R_{4/2}\sim{2.3}$ begins to emerge.
In heavier nuclei this peak remains but is now accompanied by 
nuclei with larger $R_{4/2}$ values and, in particular, a
sharp enhancement near $R_{4/2}\sim{3.33}$. Fig 3a shows the
composite distribution of $R_{4/2}$ values for all $N_B$. The peak
at $R_{4/2}\sim{2.3}$ persists.

Is this preference a fundamental feature of nuclear spectra or a
result of some bias in the data? To study this we consider in
Fig. 2(middle)
a different cut through the available $R_{4/2}$ data, namely,
distributions for different ranges of boson numbers
(i.e., valence nucleon numbers: $N_{val}=2 N_B$).
Pre-collective nuclei ($N_B\leq{3}$) show few $R_{4/2}$ values
above 2.0.  Nuclei with $N_B\geq{13}$ are nearly all rotational,
and those with $N_B$ in the range of 4-12 are
transitional. Surprisingly, no range of valence nucleon number
shows an enhancement at $R_{4/2}\sim 2.0$, contrary to the
results of the random ensembles. (In this context we recall that
the slope of 2.0 in Fig. 1 does not imply $R_{4/2}=2.0$ due to
the finite intercept of this segment.)
However, the second panel of Fig. 2(middle) shows that the peak
centered on $R_{4/2}=2.3$, mentioned above, is 
specifically associated with nuclei with $8-18$ valence
nucleons. To our knowledge, this striking
feature has not been pointed out explicitly before but, as we
shall see, leads to an empirical constraint on interactions that
is not evident from Fig. 1 alone.

Part of the explanation of the shell-by-shell results in Fig. 2(left)
therefore merely reflects changing shell sizes and hence the
distribution of possible $N_B$ values. However, this is not
sufficient.  It does not, for example, explain the overwhelming
preference for rotational $R_{4/2}$ values in regions like
$82<Z<102$ where nuclei with small numbers of valence nucleons
also exist, or the abundance of $R_{4/2}$ values near 2.3 in the
$50<Z<82$ shell. The data also reflect $\it{which}$ nuclei are
$\it{known}$ in each region. For example, 
most known nuclei with $Z>82$ have large $N_B$. 

If we assume the universality of the empirical distributions in
Fig. 2(middle), we can construct expected
$R_{4/2}$ distributions for any region of nuclei simply by
tallying the number of occurrences of a given $N_B$ for a range of
N and Z values and multiplying that number by the relevant
empirical $R_{4/2}$ distribution.  In Fig 3b, we show the
$R_{4/2}$ distribution for all nuclei with Z$<$82  assuming that, for each
proton shell, all $R_{4/2}$ values are known for neutrons filling
both the same shell and the next shell (regardless of where the
drip lines are). So, for example, for $8\leq Z<20$, we compile the
$R_{4/2}$ distribution for $8\leq N<28$. For $28\leq Z<50$, we use
$28\leq N<82$. The peak at $R_{4/2}\sim{2.3}$ persists, reflecting
simply the fact that boson numbers from 4-9 appear frequently in
virtually all shells.  Fig. 3c shows similar results for $82\leq Z < 126$ and
$82\leq N<184$. The peak at $R_{4/2}\sim 2.3$ remains, but with
a proliferation of larger boson numbers, a 
large abundance of rotational $R_{4/2}$ values near 3.33 also appears.

It is an important component of this Letter to see if calculations
with random interactions can reproduce these $R_{4/2}$
distributions.  Such calculations (for both $H_{IBM}$ and
$H_{\epsilon\kappa}$) are shown in Fig. 2(right) for comparison with the
data in Fig. 2(middle). Some results agree with the empirical trends such
as the lack of structure for low boson numbers with
$H_{IBM}$ and the abundance of $R_{4/2}\sim$3.33 values
for large $N_B$.  However, these random calculations tend to give a
peak at $R_{4/2}\sim$2 and nowhere do they show a peak at 2.3.

We will see momentarily that this points towards the need for
specific non-random interactions.  First, though, it is important
to note that, although the interactions used in $H_{\epsilon\kappa}$
are random, they actually contain an implicit bias in $R_{4/2}$
values. For $N_B=7$ and $\epsilon/\kappa$ values from $\infty$
down to $\sim{50}$, IBM spectra are vibrational. For
$\epsilon/\kappa\leq{20}$  rotational spectra result.  Thus, for a
random interaction, which effectively samples an infinitely large
range of $\epsilon/\kappa$ values, the measures of the vibrational
and rotational regions far exceed that of the transitional region,
which accounts for the high frequencies of calculated $R_{4/2}$
values near 2.0 and 3.33.  (Similar arguments apply, but for
shifted ranges of $\epsilon/\kappa$, for the other $N_B$ values.)
The Gaussian random numbers also enhance the O(6) values of
$R_{4/2}$ ($\sim 2.5$) for small $N_B$, as seen in Fig. 2 (right
column, dashes).

We carried out a large sampling of calculations to determine the
specific range of interactions that
yield a given $R_{4/2}$ value and, in particular,
$R_{4/2}\sim{2.3}$ for $N_B=4-9$.  Figure 4 shows the
statistical ranges of $\epsilon/\kappa$ and $\chi$ values that
give $R_{4/2}$ in the ranges of 2.2-2.4 and 2.8 - 3.0 for
$N_B$=7. Given the discussion just above, it is not surprising
that $R_{4/2}\sim 2.3$ values require intermediate ranges of
$\epsilon/\kappa$ near 20-30.

Thus, the empirical preference for $R_{4/2}\sim{2.3}$ values
reflects $\it{two\ equally\ necessary}$ ingredients: the relevant
interactions, and the $\it{locus}$ of accessible nuclei.  Had the
valley of stability lain elsewhere, or had the $\epsilon/\kappa$
and $\chi$ values been different, the $R_{4/2}$ distribution would
have been different.
Comparison of measured $R_{4/2}$ distributions in new, unmapped
regions with expectations (as in Figs. 3(b)-(c)) can thus be
used to deduce information either on shell structure (e.g., magic
numbers) or on the interactions applicable to these new nuclei.

Another example of characteristic nuclear data that
constrains the interaction comes from rotational nuclei, namely,
the energies of gamma vibrational modes, $E(2_2^+)$, in deformed nuclei. In
Fig. 5a we show the results for 
$H_{IBM}$ with $N_B=16$ (to approximate the mean boson number for
the deformed rare earth nuclei).  There is no evidence for a peak
in $R_{2/2}{\equiv}E(2^+_2)/E(2^+_1)$ at any value larger than
2. In heavy nuclei, however, 
such excitations are well known to occur at 10-20 times the
$2^+_1$ energy. Fig. 5b shows a collection of empirical results
for $R_{2/2}$ for the rare earth region
showing a broad bump centered on $R_{2/2}\sim{15}$.  If we now
truncate the interaction space to that relevant for deformed
nuclei, that is,  $\epsilon/\kappa<26$ we see, as shown in Fig.
5c, that, for each $\chi$ value, a distinct peak in $R_{2/2}$
probability appears.  The peak for $\chi=-0.4$ (solid), which is the
traditional value for deformed nuclei, gives an excellent
reproduction of the data. Once again, therefore, specific data on
actual nuclei help identify and constrain the nature and relative
importance of the 1- and 2-body residual interactions.

To summarize, we presented a confrontation of the data for Z=8-100 with the
predictions of nuclear structure calculations with random
interactions, in the framework of the  IBM. {\it (a)}
 General features of nuclear spectra, such as $0^+$
ground states and the appearance of vibrational and rotational nuclei\cite{4},
and the global $E(4^+_1)\ vs.\ E(2^+_1)$ trajectory, are typical
of calculations with random interactions; {\it (b)}
Heretofore unrecognized features of the data such as
the {\it absence} of an abundance peak at the vibrator value
$R_{4/2}$ = 2.0 and a frequency enhancement at $R_{4/2}\sim 2.3$,
 in nuclei with intermediate numbers of valence nucleons, were discovered.
{\it (c)} Such robust features of nuclear data, or others such as
 the characteristic
energies of $\gamma$-vibrational modes in deformed nuclei,
 point to specific ranges of interactions. In particular, we
 showed that data such as
these depend $\it{both}$ on the relevant valence nucleon
numbers (i.e., the nuclei that are known) and on the interactions.
If the residual interactions change in new nuclei (e.g., near the
drip lines), then so will the distributions of $R_{4/2}$ values
and the collective features of exotic nuclei.  If the locus of
newly accessible nuclei favors certain shell regions, the
distribution of $R_{4/2}$ values and the nature of collective
modes will reflect this as well.  Conversely, valence nucleon
number, and hence shell structure and the locations of magic
nuclei, can be estimated from empirical knowledge of $R_{4/2}$
values, even when the magic nuclei themselves are not accessible.
The usefulness of this result in studies of exotic nuclei, where
issues of shell structure or the nature of residual interactions
are paramount, is obvious.

We are grateful to Alejandro Frank, Roelof Bijker, Stuart Pittel
and Franco Iachello for discussions that initiated and 
motivated this work.  Work supported by the U.S. DOE under
Grant numbers DE-FG02-91ER-40609, DE-FG02-91ER-40608, and
DE-FG02-88ER40417.

\begin{figure}
  \begin{center}
    \epsfxsize=8.6cm\epsfbox{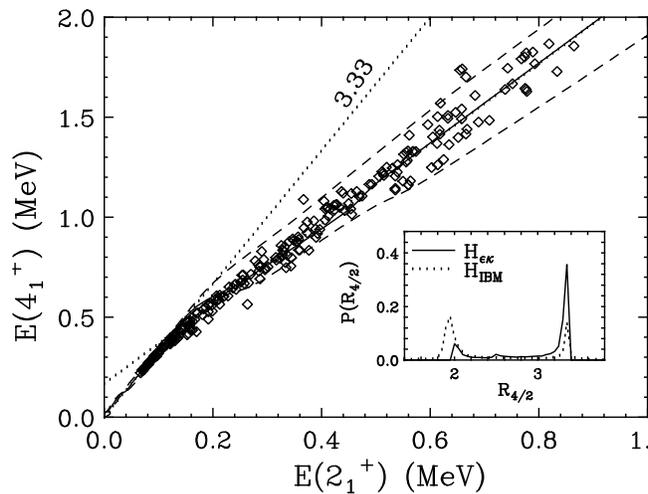}
    \caption{Plot of $E(4_1^+)$ against 
     $E(2_1^+)$ for collective nuclei with $38<Z<82$
     ($\diamond$), similar to [9].  In this and
     the following figures, we use all available data from
     [11]. The
     data lie along lines with slopes 2.0 and 3.33 (dots). The
     finite intercept of the former implies
     $E(4_1^+)=E(2_1^+)+\epsilon$. Hence
     $R_{4/2}=2+\epsilon/E(2_1^+)$ varies along this segment. 
     The statistical distribution of $E(4_1^+)$ and
     $E(2_1^+)$ energies for the full $H_{IBM}$ is a strongly
     correlated function, whose maximum is indicated by the solid
     line, and the approximate FWHM by the dashes, which agrees
     well with the data. The  variance of the Gaussian random
     numbers (which scales both axes) is adjusted to match the
     transition point between the two empirical slopes. (Inset)
     Statistical distribution of 
     $R_{4/2}$ for the full Hamiltonian (dots) and $H_{\epsilon\kappa}$
     (solid) showing peaks at 2.0 and 3.33. (In the main figure, the points
     corresponding to the abundance peak at $R_{4/2}\sim 2$ are
     at high $E(2_1^+)$, offscale at the upper right.)
     All simulations are for $N_B=16$ and 
     50000 realizations of the random Hamiltonians.}
    \label{fig:thr}
  \end{center}
\end{figure}

\begin{figure} 
\begin{center}
    \leavevmode
  \epsfysize=6.5cm\epsfbox{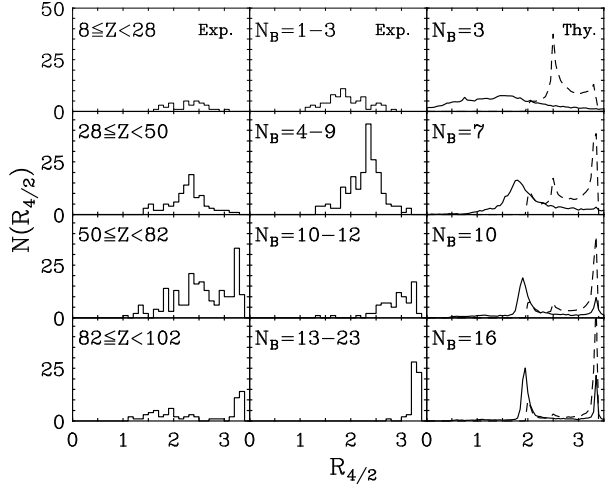}
  \caption{(Left column) Experimental distribution of
  $R_{4/2}$ as a function of proton shells. (Middle column)
  Experimental distribution of 
  $R_{4/2}$ as a function of $N_B$. (The number of valence
  nucleons is $2N_B$.)  (Right column) Statistical
  predictions of  $R_{4/2}$ for
  selected boson numbers using $H_{IBM}$ (solid) and $H_{\epsilon
  \kappa}$ (dashes) with random $\chi$. The vertical scale for this
  column is arbitrary.}
  \label{fig:one}
  \end{center}
\end{figure}

\begin{figure}
  \begin{center}
    \epsfxsize=8.5cm\epsfbox{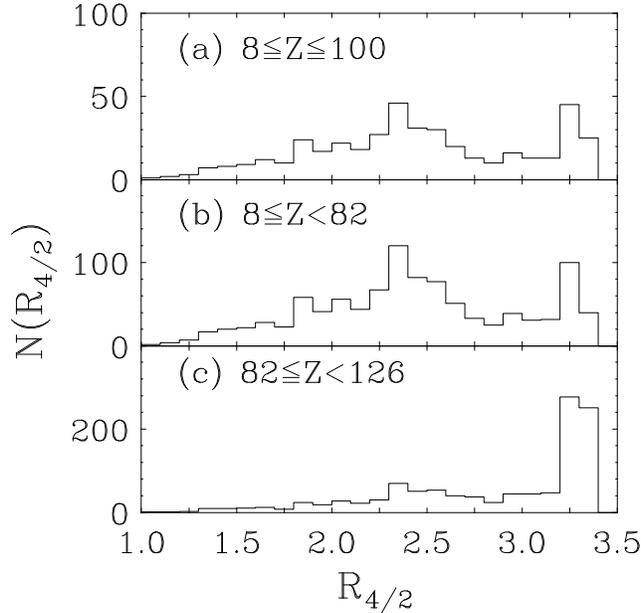}
    \caption{Empirical distributions of $R_{4/2}$ for
      indicated proton regions. (a) Sum of existing data from
      Fig. 2(left). 
      (b) and (c) Semi-empirical estimates of $R_{4/2}$ distributions for
      large regions of nuclei where it is assumed that the $R_{4/2}$
      values would be available for full proton and neutron shells.
      Specifically, for each full proton shell, we take the set of nuclei
      obtained when the neutrons fill the same and next higher shell.
      These panels are constructed by assuming that the empirical distributions 
      in Fig. 2 (middle) apply to all nuclei and, therefore, by folding these 
      distributions with the frequency of $N_B$ for these regions.}
    \label{fig:two}
  \end{center}
\end{figure}
 
\begin{figure}
  \begin{center}
    \epsfxsize=8.8cm\epsfbox{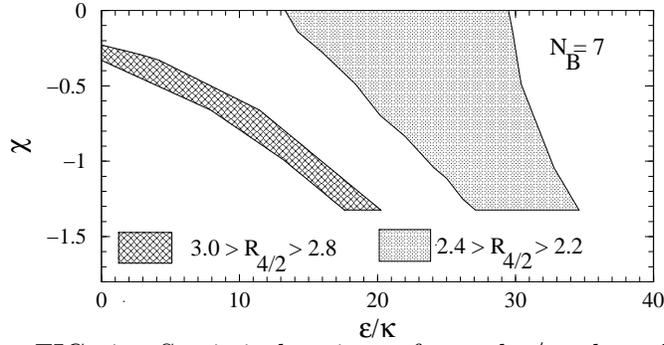}
    \caption{ Statistical regions of $\chi$ and
  $\epsilon/\kappa$ where $R_{4/2}$ falls in the
  range 2.2--2.4 and 2.8--3.0 for $N_B=7$.}
    \label{fig:fou}
  \end{center}
\end{figure}
\begin{figure} 
  \begin{center}
    \epsfxsize=8.cm\epsfbox{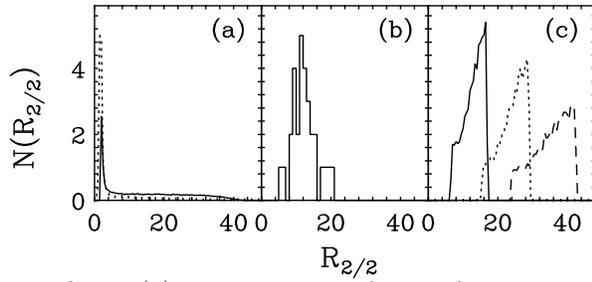}
    \caption{(a) Distribution of $R_{2/2}$ for $H_{IBM}$
     (dots) and $H_{\epsilon\kappa}$ with random $\chi$ (solid). (b) 
     Experimental
     distribution of $R_{2/2}$. (c) Distributions for  $H_{\epsilon
     \kappa}$ for $\epsilon/\kappa<26$ and $\chi=-\sqrt{7}/2$ (dashes),
     -0.8 (dots) and -0.4 (solid), showing the selectivity in
     $R_{2/2}$. }
    \label{fig:fiv}
  \end{center}
\end{figure}

\end{document}